\documentclass[reprint%
,aip%
,pop%
,longbibliography%
]{revtex4-2}

\usepackage[english]{babel}
\usepackage{amsmath}
\usepackage{amssymb}
\usepackage{graphicx}
\usepackage{dcolumn}
\usepackage{tikz}
\usepackage{bm}
\usepackage{xcolor}
\usepackage{xpatch}
\usepackage{natbib}

\newcommand\mi{\mathrm i}
\newcommand\axion{\phi}
\newcommand\gagg{g_{a\gamma\gamma}}

\newcommand{\nold}{^{n}}
\newcommand{\nhalf}{^{n+\frac{1}{2}}}
\newcommand{\nnew}{^{n+1}}
\newcommand{\colorhighlight}[1]{\textcolor{black}{#1}}
\newcommand{\dt}{\Delta t}
\newcommand{\dx}{\Delta x}
\newcommand{\pga}{P_{\gamma\rightarrow\axion}}
\newcommand{\pag}{P_{\axion\rightarrow\gamma}}
\newcommand{\dd}{\mathop{}\!\mathrm{d}}
\newcommand{\DD}{\mathop{}\!\mathrm{D}}

\makeatletter
\let\ORIbbl@fixname\bbl@fixname
\def\bbl@fixname#1{%
  \@ifundefined{languagealias@\expandafter\string#1}
    {\ORIbbl@fixname#1}
    {\edef\languagename{\@nameuse{languagealias@#1}}}%
}
\newcommand{\definelanguagealias}[2]{%
  \@namedef{languagealias@#1}{#2}%
}
\makeatother

\definelanguagealias{en}{english}
\definelanguagealias{en-US}{english}
\definelanguagealias{zh}{chinese}

\begin{document}

\title{Modeling of axion and electromagnetic fields coupling in a particle-in-cell code}

\author{Xiangyan An}
\affiliation{Key Laboratory for Laser Plasmas (MoE), School of Physics and Astronomy, Shanghai Jiao Tong University, Shanghai 200240, China}
\affiliation{Collaborative Innovation Center of IFSA, Shanghai Jiao Tong University, Shanghai 200240, China}
\affiliation{Tsung-Dao Lee Institute, Shanghai Jiao Tong University, Shanghai 200240, China}

\author{Min Chen}
\email{minchen@sjtu.edu.cn}
\affiliation{Key Laboratory for Laser Plasmas (MoE), School of Physics and Astronomy, Shanghai Jiao Tong University, Shanghai 200240, China}
\affiliation{Collaborative Innovation Center of IFSA, Shanghai Jiao Tong University, Shanghai 200240, China}

\author{Jianglai Liu}
\affiliation{Tsung-Dao Lee Institute, Shanghai Jiao Tong University, Shanghai 200240, China}

\author{Zhengming Sheng}
\affiliation{Key Laboratory for Laser Plasmas (MoE), School of Physics and Astronomy, Shanghai Jiao Tong University, Shanghai 200240, China}
\affiliation{Collaborative Innovation Center of IFSA, Shanghai Jiao Tong University, Shanghai 200240, China}
\affiliation{Tsung-Dao Lee Institute, Shanghai Jiao Tong University, Shanghai 200240, China}

\author{Jie Zhang}
\affiliation{Key Laboratory for Laser Plasmas (MoE), School of Physics and Astronomy, Shanghai Jiao Tong University, Shanghai 200240, China}
\affiliation{Collaborative Innovation Center of IFSA, Shanghai Jiao Tong University, Shanghai 200240, China}
\affiliation{Tsung-Dao Lee Institute, Shanghai Jiao Tong University, Shanghai 200240, China}

\date{\today}

\begin{abstract}
	Axions have aroused widespread research interest because they can solve the strong CP problem and serve as a possible candidate for dark matter. Currently, people have explored a lot of axion detection experiments, including passively detecting the existing axions in the universe, and actively generating axions in the laboratory. Recently, axion-coupled laser-plasma interactions have been discussed as a novel method to detect axions. Petawatt (PW) lasers are considered as a powerful tool to study not only the vacuum polarization but also the axion coupling, due to their extreme fields. However, particle-in-cell (PIC) simulation is still missed in current studies, which limits the understanding of axion-coupled laser-plasma interactions. In this paper, we proposed the method to include the axion field and the coupling with electromagnetic (EM) fields in PIC codes. The axion wave equation and modified Maxwell's equations are numerically solved, while the EM field modulation from axions is considered as a first-order perturbation. Meanwhile, different axion field boundary conditions are considered to satisfy different simulation scenarios. The processes of conversions between axions and photons, and weak laser pulse propagation with axion effects are checked as benchmarks of the code. Such an extended PIC code may help researchers develop novel axion detection schemes based on laser-plasma interactions and provide a better understanding of axion-coupled astrophysical processes.
\end{abstract}

\maketitle

%%%% Start %%%%%%
\section{Introduction}
\label{sec-introduction}

Axion was initially proposed by Wilczek~\cite{wilczek_problem_1978} and Weinberg~\cite{weinberg_new_1978} as the resulted particle from spontaneously broken of Peccei-Quinn (PQ) symmetry~\cite{peccei_cp_1977,peccei_constraints_1977}, to solve the strong CP problem~\cite{kim_axions_2010}. Besides, axions are also considered as one of the possible candidates for dark matter~\cite{kim_axions_2010,chadha-day_axion_2022}.

Currently, there are many axion detection experiments around the world. One way is to detect the existing axions, including those from the sun and universe background. In this scenario, there are Axion Dark Matter Experiment (ADMX)~\cite{du_search_2018,braine_extended_2020,bartram_search_2021} to convert dark matter axions into microwave photons and CERN Axion Solar Telescope (CAST)~\cite{cast_collaboration_new_2017} to convert solar axions into X-ray photons in magnetic fields to detect axions. Besides, people are also seeking axion hints in some astrophysical phenomena~\cite{foster_green_2020,edwards_transient_2021,buschmann_axion_2021}.

Besides passively detecting the existing axions in the universe, some groups also actively generate axions and detect the derivative effects. Polarization rotation of a laser in a transverse magnetic field is thought to be a signal of axion existence, such as PVLAS~\cite{zavattini_experimental_2006}, BMV~\cite{battesti_bmv_2008}, BFRT~\cite{cameron_search_1993}, and Q\&A~\cite{chen_q_2007} experiments. Meanwhile, there is also an effective laser experiment named "light shining through a wall". A laser is used to generate axions through a magnetic region, after which a wall is used to block the laser photon and only the axions can transmit the wall. Then the axions will be reconverted into photons in a second magnetic region to be detected. Such a scheme is used in ALPS~\cite{ehret_new_2010}, GammeV~\cite{chou_search_2008}, LIPSS~\cite{afanasev_experimental_2008}, OSQAR~\cite{pugnat_results_2008}, and BMV~\cite{robilliard_no_2007}.

Although there are currently many detection and experimental setups to search axions and other dark matters in the world, no convincing signals have been discovered so far~\cite{li_search_2023}. One possible reason is the weak coupling between axions and photons, $\gagg\lesssim 10^{-10}\,\text{GeV}^{-1}$. Meanwhile, high-power lasers are recently considered to enhance the vacuum birefringence (VB) signal~\cite{dinu_vacuum_2014,shen_exploring_2018,dai_fermionic_2024}, which is also the aim of PVLAS, BMV, BFRT and Q\&A experiments. Instead of polarization rotation from axion coupling, an initially linear-polarized laser may gain a small ellipticity after propagating through the magnetic field. Such an ellipticity signal can be significantly enhanced with a 10\,PW laser (with magnetic fields reaching $10^6$ T). Similarly, the polarization rotation signal may also be enhanced through PW lasers.

Moreover, when the laser intensity increases, there could be more new axion-related phenomena to study and detect in laser-plasma interactions. People have found a new quasiparticle from axions coupling to the electrostatic (Langmuir) modes~\cite{tercas_axion-plasmon_2018}. Some plasma-based axion generation and detection schemes have also been proposed and discussed~\cite{mendonca_axion_2020,mendonca_axion_2020-1,huang_axion-like_2022}.

However, current studies about axion-coupled laser-plasma interactions are still limited to theoretical calculations. To get a comprehensive understanding of the interaction process, PIC is a quite powerful tool. In a PIC code, particle motion is calculated according to the Lorentz force with modifications from some QED processes, and EM fields are numerically solved according to Maxwell's equations and the current contributed by the particles. Such a method has been successfully used in studying laser wakefield acceleration~\cite{mangles_monoenergetic_2004,geddes_high-quality_2004,faure_laserplasma_2004,an_bragg_2022}, strong field quantum electrodynamics (QED) effects~\cite{gonoskov_extended_2015,song_dense_2022}, and so on. Despite that, the axion field and the coupling with EM fields are still missed in current usually used PIC codes, which limited the studies about axion-coupled laser-plasma interactions.

In this paper, we added the axion field and the coupling with EM fields into the PIC code EPOCH~\cite{arber_contemporary_2015}. The axion wave equation and modified Maxwell's equations are numerically solved. Since the axion coupling is extremely small, the axion modulation to EM fields is considered a first-order perturbation. Meanwhile, different kinds of axion field boundary conditions are considered to satisfy different simulation scenarios. With such modifications, the code can simulate axion fields self-consistently. In the end, the axion generation from photons and conversion into photons in a constant magnetic field are shown as benchmarks of the code.
%Such a weak coupling would result in a sufficiently low conversion rate of both axions into photons and photons into axions. As we know, the conversion probability between axions and photons in a constant magnetic field $B$ is proportional to the factor $\left(\gagg Bl\right)^2$, where $l$ is the propagating distance. Thus using a stronger magnetic field may significantly improve the conversion rate. Laser-plasma interaction has recently been found to produce $10^{6}\,\text{T}$ magnetic field in $10\,\mu\text{m}$ distance~\cite{xue_generation_2023},

%Axion is generally interesting. There are many types of research, including search experiments (dark matter, sun, astronomy) and mechanism studies.

% High-power laser may be a new method. To self-consistently study, we need particle-in-cell simulation. PIC is ....

%We have included the axion field and the coupling with EM fields in PIC code EPOCH~\cite{arber_contemporary_2015}. In this paper,...

\section{Simulation Methods}
\label{sec-implementation}

\subsection{Units and Field Equations}
%We have implemented axion fields and the coupling with electromagnetic (EM) fields in the PIC code EPOCH~\cite{arber_contemporary_2015}. In this section, we will give the model to describe axion fields in the code.

In natural units ($\hbar=\varepsilon_0=c=1$), the axion coupling with EM fields can be described by the Lagrangian density term $\mathcal{L}_\text{int}=  -\frac{1}{4}\gagg \axion F_{\mu\nu}G^{\mu\nu}=\gagg \axion\bm E\cdot\bm B$, where $\gagg$ is the coupling constant, $F^{\mu\nu}$ and $G^{\mu\nu}$ are the electromagnetic tensor and dual tensor, respectively, and $\bm E$, $\bm B$ are the electric field and magnetic field, respectively. The total Lagrangian density reads,
\begin{gather}
	\mathcal{L} = \mathcal{L}_\text{EM}+\mathcal{L}_{\axion}+\mathcal{L}_\text{int}, \\
	\mathcal{L}_\text{EM} = -\frac{1}{4}F_{\mu\nu}F^{\mu\nu}+A_\mu j^\mu_e, \\
	\mathcal{L}_{\axion} = \frac{1}{2}\partial_\mu \axion\partial^\mu \axion-\frac{1}{2}m^2_a\axion^2,
\end{gather}
where $m_a$ is the axion mass, $\mathcal{L}_\text{EM}$ and $\mathcal{L}_{\axion}$ are the Lagrangian density of EM field and free axion field, respectively. After variation to the total Lagrangian density, we can get the modified Maxwell equations with axion coupling,
\begin{gather}
	\partial_t^2 \axion-\nabla ^{2} \axion+m^{2}_a \axion =g_{a\gamma \gamma }\ \bm {E} \cdot \bm {B}, \label{eq:wave_axion}\\
	{\displaystyle \nabla \cdot \bm {E} =\rho -g_{a\gamma \gamma }\bm {B} \cdot \nabla \axion}, \label{eq:divergence-e-axion}\\
	{\displaystyle \nabla \cdot \bm {B} =0}, \\
	{\displaystyle \nabla \times \bm {E} =-\partial_t\bm B}, \\
	{\displaystyle \nabla \times \bm {B} ={\partial_t\bm E}+\bm{j} +g_{a\gamma \gamma }[(\partial_t \axion)\bm {B} -\bm {E} \times \nabla \axion]},
\end{gather}
where $\rho$ and $\bm j$ are the electrical charge density and current density, respectively.
Here we did not introduce an additional duality symmetry as some other works~\cite{visinelli_axion-electromagnetic_2013,tercas_axion-plasmon_2018,mendonca_axion_2020}.

The code EPOCH is an open source PIC code and it's widely used in laser plasma community~\cite{ridgers_dense_2012,charpin_simulation_2024,arber_contemporary_2015,vyskocil_simulations_2018}. EPOCH makes calculations in SI units. And in SI units, we choose the axion field $\axion$ to have dimensions of frequency, and the axion coupling constant $\gagg$ to have dimensions of time. In this way, the modified Maxwell equations in SI units read,
\begin{gather}
	\left(\frac{\partial^2}{c^2\partial t^2}-\nabla^2+\frac{m^2_ac^2}{\hbar^2}\right)\axion=\frac{\gagg }{\hbar\mu_0}\bm E\cdot\bm B, \\
	{\displaystyle \nabla \cdot \bm {E} =\frac{\rho}{\varepsilon_0} -cg_{a\gamma \gamma }\bm {B} \cdot \nabla \axion}, \\
	{\displaystyle \nabla \cdot \bm {B} =0}, \\
	{\displaystyle \nabla \times \bm {E} =-\partial_t\bm B}, \\
	{\displaystyle \nabla \times \bm {B} =\frac{\partial}{c^2\partial t}\bm E+\mu_0\bm j +\frac{\gagg}{c}[(\partial_t \axion)\bm {B} -\bm {E} \times \nabla \axion]}.
\end{gather}

\subsection{Perturbations due to Axion Field\label{sec:perturbation}}
First of all, we need to notice that the axion modulation to the EM fields might be so small that it would be overwhelmed by the floating point error. To solve this problem, we use field perturbation separation (FPS) method by expanding the variables according to the order of axion coupling: $\bm E=\bm E_0+\bm E_1,\,\bm B=\bm B_0+\bm B_1,\,\rho = \rho_0+\rho_1,\,\bm j=\bm j_0+\bm j_1$, where $\bm E_0,\,\bm B_0,\,\rho_0,\,\bm j_0$ satisfy the Maxwell equations without axion. Then the first-order perturbations of the EM fields satisfy,
\begin{gather}
	\nabla \cdot \bm{E}_1 =\frac{\rho_1}{\varepsilon_0} -cg_{a\gamma \gamma }\bm{B}_0 \cdot \nabla \axion, \label{eq:divergence-e1}\\
	\nabla \cdot \bm{B}_1 =0, \\
	\nabla \times \bm{E}_1 =-\partial_t\bm B_1, \\
	\nabla \times \bm{B}_1 =\frac{\partial}{c^2\partial t}\bm E_1+\mu_0\bm j_1 +\frac{\gagg}{c}[(\partial_t \axion)\bm{B}_0 -\bm{E}_0 \times \nabla \axion].\label{eq:curl-b1}
\end{gather}

As an estimation, for a laser with the normalized vector potential $a_0=\left|e E_0/(m_e\omega_0c)\right|=100$ (where $\omega_0=ck_0=2\pi c/\lambda_0$ is the laser frequency, and $e,\,m_e$ are the electron charge and mass, respectively), the non-perturbed electric field is $E_0 = 4\times 10^{14}\,\text{V/m}$ for an 800\,nm laser. Considering such a laser generating axions in a magnetic field as strong as itself within a wavelength, the generated axion field is approximately $\frac{\gagg E_0B_0}{2\mu_0\hbar k_0^2}=5.7\times 10^{9}\,\text{s}^{-1}$ (for a typical $\gagg=2.7\times 10^{-13}\hbar\,\text{GeV}^{-1}$). With such an axion field, the axion modulation to the EM field can be estimated as $E_1\approx\gagg\axion B_0c =4\times 10^{-13}\,\text{V/m}$.

Unfortunately, the value of $ E_0$ is represented as a double precision floating-point number in the program. In binary, such a number is $E_0\stackrel{\text{bin}}{=}1.\underbrace{011\cdots0}_{52}\times 2^{48}\,\text{V/m}$. To represent such a number, the computer would use 1 bit for the $\pm$ sign, 11 bits for the exponent, and 52 bits for the mantissa. Therefore, the minimal step for $E_0$ to increase is $\mathop\text{error}\left(E_0\right)=2^{-52}\times 2^{48}=0.0625\,\text{V/m}$, which is called the floating point error of $E_0$. We can see that $E_1$ is even far less than the floating point error: $E_1\ll \mathop\text{error}\left(E_0\right)$.  Therefore, it is necessary to treat the zero and first order EM fields independently according to Eq.~\ref{eq:divergence-e1}-\ref{eq:curl-b1}. Otherwise, $E_1$ would be overwhelmed by the floating point error of $E_0$ and we can never get the correct modulation from the axion fields.

Moreover, since Eq.~\ref{eq:divergence-e1} can be derived from Eq.~\ref{eq:curl-b1} and the continuity equation $\partial_t\rho_1+\nabla\cdot\bm j_1=0$ together, we only need to treat the current perturbation $\bm j_1$ in the code. For this part, we consider the perturbation to the particle momentum and velocity: $\bm p=\bm p_0+\bm p_1,\,\bm v=\bm v_0+\bm v_1$. To calculate the perturbed particle motion, we should first calculate the perturbed fields felt by the particle, which are given by ($F=E_{x,y,z},\,B_{x,y,z}$ is the example field component):
\begin{gather}
	\begin{split}
		\left.F\right|_{\bm x=\bm x_0+\bm x_1} & =\left.\left(F_0+F_1\right)\right|_{\bm x=\bm x_0+\bm x_1}                                                                      \\
		                                       & =\sum_{ijk}S(\bm r_{ijk}-\bm x_0-\bm x_1)\left(F_{0,ijk}+F_{1,ijk}\right)                                                       \\
		                                       & \approx\sum_{ijk}S(\bm r_{ijk}-\bm x_0)F_{0,ijk}                                                                                      \\
		                                       & +\sum_{ijk}\left[\left.\frac{\partial S(\bm r_{ijk}-\bm x)}{\partial \bm x}\right|_{\bm x=\bm x_0}\cdot\bm x_1\right] F_{0,ijk} \\
		                                       & +\sum_{ijk}S(\bm r_{ijk}-\bm x_0)F_{1,ijk}                                                                                      \\
		                                       & =F_\text{nop}+F_\text{p},
	\end{split}
\end{gather}
where $S(\bm r_{ijk}-\bm x)$ is the shape function of a particle locating at position $\bm x$, $\bm r_{ijk}$ is the space grid position of the cooresponding field component, $F_\text{nop}=\sum_{ijk}S(\bm r_{ijk}-\bm x_0)F_{0,ijk}$ and $F_\text{p}$ are the non-perturbed and perturbed fields felt by the particle. The the perturbed position and momentum of the particle can be calculated according to:
\begin{gather}
	\frac{\dd \bm x_1}{\dd t}=\bm v_1, \\
	\frac{\dd \bm p_1}{\dd t} = q\left(\bm E_\text{p}+\bm v_0\times \bm B_\text{p}+\bm v_1\times\bm B_\text{nop}\right), \\
	\bm v_1 =\frac{\partial \bm v}{\partial\bm p}\cdot\bm p_1= \frac{c}{\gamma_0^3}\left[\gamma_0^2\bm u_1-\bm u_0\left(\bm u_0\cdot\bm u_1\right)\right],
\end{gather}
where $q$ and $m$ are the charge and mass of the particle, respectively, and $\bm u={\bm p}/\left({mc}\right)$ is the normalized momentum. Then the current perturbation can be calculated according to the particle shape function:
\begin{gather}
	\begin{split}
		\left(\bm j_0+\bm j_1\right)_{ijk} & =\frac{q}{\Delta V}\left[ S(\bm r_{ijk}-\bm x_0-\bm x_1)\left(\bm v_0+\bm v_1\right)\right]                                                                                                                                                   \\
										   & \approx\frac{q}{\Delta V}S(\bm r_{ijk}-\bm x_0)\bm v_0 \phantom{=}\\
										   &+\frac{q}{\Delta V}\left[\left.\frac{\partial S(\bm r_{ijk}-\bm x)}{\partial \bm x}\right|_{\bm x=\bm x_0}\cdot\bm x_1\right] \bm v_{0}\\
										   &+\frac{q}{\Delta V}S(\bm r_{ijk}-\bm x_0)\bm v_{1},
	\end{split}
\end{gather}
where $\Delta V$ is the macroparticle volume in the code. So far we have given the full description of the perturbed EM fields in the code, which is necessary to correctly calculate the derivative effects of axions.

\subsection{Field Equations Discretization}
In the following, we give the description of axion discretization in both time and space grids. The superscription will label the index of time steps, and the subscription will label the space index. It should be pointed out that a single 0 or 1 on the subscript is used to label the order of the EM field. Although we use SI units in the code, the following equations used in the numerical model will still be shown in natural units for convenience. From the equations, we naturally set the axion fields at the half-time grids. Then the time differential is realized as,
\begin{gather}
	\partial_t\axion \rightarrow \left(\DD_t \axion\right)^n = \frac{\axion^{n+\frac{1}{2}}-\axion^{n-\frac{1}{2}}}{\Delta t}, \\
	\partial_t^2\axion \rightarrow \left(\DD_t^2\axion\right)^{n+\frac{1}{2}} = \frac{\axion^{n+\frac{3}{2}}-2\axion^{n+\frac{1}{2}}+\axion^{n-\frac{1}{2}}}{\left(\Delta t\right)^2},
\end{gather}
where $\Delta t$ is the time step. In one time step of the code, we would first update the EM fields for half time step and then update the axion field,
\begin{gather}
	\begin{split}
		{\bm E}\nhalf_1 = {\bm E}\nold_1 & + \frac{\Delta t}{2} \Bigl(\nabla\times {\bm B}\nold_1 - {\bm j}\nold_1                             \\
		                                 & -{\gagg}\left[\left(\DD_t \axion\right)^{n}{\bm B}^n_0-\bm E^n_0\times\nabla \axion^n\right]\Bigr),
	\end{split}\label{eq:update_E}\\
	{\bm B}\nhalf_1 = {\bm B}\nold_1 - \frac{\Delta t}{2}\nabla \times {\bm E}\nhalf_1, \\
	\frac{\axion^{n+\frac{3}{2}}-2\axion^{n+\frac{1}{2}}+\axion^{n-\frac{1}{2}}}{\left(\Delta t\right)^2}=\left(\nabla^2 \axion-m^2_a\axion +\gagg \bm E_0\cdot\bm B_0\right)^{n+\frac{1}{2}}. \label{eq:update_axion}
\end{gather}
	The particles are then pushed by both zero and first order fields. After that, we update the EM fields for the left half time step.
	\begin{gather}
	{\bm B}\nnew_1 = {\bm B}\nhalf_1 - \frac{\Delta t}{2}\nabla \times {\bm E}\nhalf_1, \\
	\begin{split}
		{\bm E}\nnew_1 = {\bm E}\nhalf_1 & + \frac{\Delta t}{2}\Bigl(\nabla\times {\bm B}\nnew_1 - {\bm j}\nnew_1                                           \\
		                                 & -{\gagg}\left[\left(\DD_t \axion\right)\nhalf{\bm B}\nnew_0-{\bm E}\nnew_0\times\nabla \axion\nnew\right]\Bigr).
	\end{split}
\end{gather}
Here $\axion^n$ would be given by the averaged axion fields at two time steps,
\begin{gather}
	\axion^n = \frac{1}{2}\left(\axion^{n+\frac{1}{2}}+\axion^{n-\frac{1}{2}}\right).
\end{gather}

\begin{figure}[htbp]
	\centering
	\includegraphics[scale=1.4]{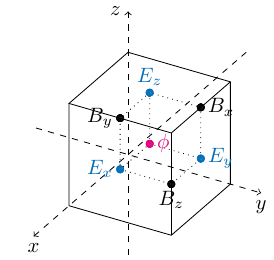}
	\caption{The configuration of EM fields and axion field in the PIC code. \label{fig:config}}
\end{figure}

As for the spatial discretization, the configuration is shown in Fig.~\ref{fig:config}. While the components of EM fields are staggered to the grid boundaries in certain directions, we consider the axion field to locate at the cell center in all directions. In this way, we can calculate the space dependent terms in Eq.~\ref{eq:update_E} and \ref{eq:update_axion}.
\begin{gather}
	i' = i+\frac{1}{2},\quad j' = j+\frac{1}{2},\quad k' = k+ \frac{1}{2}, \\
	\partial_x^2\axion\rightarrow \left(\DD_x^2\axion\right)_{i'j'k'}=\frac{\axion_{(i'+1)j'k'}-2\axion_{i'j'k'}+\axion_{(i'-1)j'k'}}{\left(\Delta x\right)^2}.
\end{gather}

Moreover, to update the axion fields as Eq.~\ref{eq:update_axion}, we need to calculate the source term $\bm E_0\cdot \bm B_0$ at axion field locations. The actual values of the EM fields are given by averaging the fields at the nearest cells,
\begin{gather}
	\left(\bm E_0\cdot\bm B_0\right)_{i'j'k'}= \bar{E}_{0x}\bar{B}_{0x}+\bar{E}_{0y}\bar{B}_{0y}+\bar{E}_{0z}\bar{B}_{0z},\\
	\bar{E}_{0x} = \frac{1}{2}\left[E_{0x,\left(i'+\frac{1}{2}\right)j'k'}+E_{0x,\left(i'-\frac{1}{2}\right)j'k'}\right],\\
	\begin{split}
		\bar{B}_{0x}=\frac{1}{4}
		\Bigl[ & B_{0x,i'\left(j'+\frac{1}{2}\right)\left(k'+\frac{1}{2}\right)}
		+B_{0x,i'\left(j'+\frac{1}{2}\right)\left(k'-\frac{1}{2}\right)}                                                                                \\
		+      & B_{0x,i'\left(j'-\frac{1}{2}\right)\left(k'+\frac{1}{2}\right)}+B_{0x,i'\left(j'-\frac{1}{2}\right)\left(k'-\frac{1}{2}\right)}\Bigr].
	\end{split}
\end{gather}
And the other components are averaged similarly. The cross term $\bm E_0\times\nabla\axion$ in Eq.~\ref{eq:update_E} is also given by averaging. Here we show the expression of one component $\left(\bm E_0\times \nabla \axion\right)_{x}$, the other components are similar.
%\begin{widetext}
	\begin{gather}
		\left(\bm E_0\times \nabla \axion\right)_{x}\rightarrow\left(E_{0y}\DD_z \axion-E_{0z}\DD_y \axion\right)_{\left(i'+\frac{1}{2}\right)j'k'}, \\
		\begin{split}
			\left(E_{0y}\DD_z \axion\right)_{\left(i'+\frac{1}{2}\right)j'k'} 
			&=  \frac{1}{4}\left[E_{0y,i'\left(j'+\frac{1}{2}\right)k'}+E_{0y,i'\left(j'-\frac{1}{2}\right)k'}\right]\\
			&\phantom{=}\cdot\frac{\axion_{i'j'\left(k'+1\right)}-\axion_{i'j'\left(k'-1\right)}}{2\Delta z}                                                              \\
			&+ \frac{1}{4}\left[E_{0y,\left(i'+1\right)\left(j'+\frac{1}{2}\right)k'}+E_{0y,\left(i'+1\right)\left(j'-\frac{1}{2}\right)k'}\right]\\
			&\phantom{=}\cdot\frac{\axion_{\left(i'+1\right)j'\left(k'+1\right)}-\axion_{\left(i'+1\right)j'\left(k'-1\right)}}{2\Delta z}.
		\end{split}
	\end{gather}
%\end{widetext}

\subsection{Axion Field Boundary Conditions}

The solutions of partial differential equations are only complete with specific boundary conditions. As for the boundary conditions of axions, we implemented four kinds of boundary conditions in the code: reflect, periodic, outflow, and PML (Perfectly Matched Layers). These are quite similar as EM field treatments in usual PIC codes, which are used for different scenarios~\cite{berenger_perfectly_1994,roden_convolution_2000,oskooi_failure_2008}.
\begin{enumerate}
	\item Reflect. As for the reflect boundary condition, we set the axion field gradient to zero at the boundary.
	      \begin{gather}
		      \axion_{\left(-\frac{1}{2}\right)j'k'}=\axion_{\left(\frac{1}{2}\right)j'k'},\quad \cdots.
	      \end{gather}
	\item Periodic. As for the periodic boundary condition, we would set the axion field at the out cell to be the same on the other side.
	      \begin{gather}
		      \axion_{\left(-\frac{1}{2}\right)j'k'} = \axion_{\left(N_x-\frac{1}{2}\right) j'k'},\quad \cdots,
	      \end{gather}
	      where $N_x$ is the total grid number in $x$ direction.
	\item Outflow. As for the outflow boundary condition, we would assume a left-handed wave propagating along $x$ direction at the $x_\text{min}$ boundary, or a right-handed wave at the $x_\text{max}$ boundary. And the treatments in the $y$ and $z$ directions are similar. Specifically, we solve the two equations at the boundary to obtain the axion fields $\axion_{\left(-\frac{1}{2}\right)j'k'}^{n}$ and $\axion_{\left(-\frac{3}{2}\right)j'k'}^{n-1}$ out of the simulation box at each time step, which are needed to update the axion field in the box. The results are (with the source term $S=\gagg \bm E_0\cdot\bm B_0$),
	      \begin{widetext}
		      %	\begin{gather}
		      %		\frac{\colorhighlight{\axion_{\left(-\frac{1}{2}\right)j'k'}^{n}}-\axion_{\left(-\frac{1}{2}\right)j'k'}^{n-2}}{2\dt}=\frac{\axion_{\left(\frac{1}{2}\right)j'k'}^{n-1}-\colorhighlight{\axion_{\left(-\frac{3}{2}\right)j'k'}^{n-1}}}{2\dx}, \\
		      %		\begin{split}
		      %        &\frac{\colorhighlight{\axion_{\left(-\frac{1}{2}\right)j'k'}^{n}}-2\axion_{\left(-\frac{1}{2}\right)j'k'}^{n-1}+\axion_{\left(-\frac{1}{2}\right)j'k'}^{n-2}}{\left(\dt\right)^2}-\frac{\axion_{\left(\frac{1}{2}\right)j'k'}^{n-1}-2\axion_{\left(-\frac{1}{2}\right)j'k'}^{n-1}+\colorhighlight{\axion_{\left(-\frac{3}{2}\right)j'k'}^{n-1}}}{\left(\dx\right)^2}\\
		      %		=&-m^2_a\axion_{\left(-\frac{1}{2}\right)j'k'}^{n-1}+\gagg \bm E_{\left(-\frac{1}{2}\right)j'k'}^{n-1}\cdot\bm B_{\left(-\frac{1}{2}\right)j'k'}^{n-1}=S_{\left(-\frac{1}{2}\right)j'k'}^{n-1}-m^2_a\axion_{\left(-\frac{1}{2}\right)j'k'}^{n-1}.
		      %		\end{split}
		      %	\end{gather}
		      %	The results are,
		      \begin{gather}
			      \begin{split}
				      \left[\frac{\dt}{\dx}+\frac{\left(\dt\right)^2}{\left(\dx\right)^2}\right]\colorhighlight{\axion_{\left(-\frac{3}{2}\right)j'k'}^{n-1}}= & -2\axion_{\left(-\frac{1}{2}\right)j'k'}^{n-1}+2\axion_{\left(-\frac{1}{2}\right)j'k'}^{n-2}-\left(\dt\right)^2 \left[S_{\left(-\frac{1}{2}\right)j'k'}^{n-1}-m_a^2\axion_{\left(-\frac{1}{2}\right)j'k'}^{n-1}\right] \\
				                                                                                                                     & +\frac{\dt}{\dx}\axion_{\left(\frac{1}{2}\right)j'k'}^{n-1}-\frac{\left(\dt\right)^2}{\left(\dx\right)^2}\left[\axion_{\left(\frac{1}{2}\right)j'k'}^{n-1}-2\axion_{\left(-\frac{1}{2}\right)j'k'}^{n-1}\right],
			      \end{split}\\
			      \colorhighlight{\axion_{\left(-\frac{1}{2}\right)j'k'}^{n}}=\axion_{\left(-\frac{1}{2}\right)j'k'}^{n-2}+\frac{\dt}{\dx}\left[\axion_{\left(\frac{1}{2}\right)j'k'}^{n-1}-\colorhighlight{\axion_{\left(-\frac{3}{2}\right)j'k'}^{n-1}}\right].
		      \end{gather}
	      \end{widetext}
	\item PML. As for the outflow boundary condition, we would assume the axion field to exponentially decay in the absorbing layer,
	      \begin{gather}
		      \axion = \axion_{\text{i}} e^{\mi(\omega t - k x)-\sigma t },
	      \end{gather}
	      where $\sigma$ is an artificial absorbing constant.
	      Thus we made the transform $\partial_t\rightarrow \partial_t+\sigma$ in Eq.~\ref{eq:wave_axion}. Then after discretization, the evolution equation for the axion field in the absorbing layer is,
	      \begin{gather}
		      \begin{split}
			      \left(1+\sigma \dt\right)\axion_{i'j'k'}^{n+1}= & \left(\dt\right)^2\bigl[S_{i'j'k'}^n-m^2\axion_{i'j'k'}^{n}                                             \\
			                                                      & +\left(\DD_x^2+\DD_y^2+\DD_z^2\right)\axion^{n}_{i'j'k'}-\sigma^2\axion^{n}_{i'j'k'}\bigr] \\
			                                                      & +\sigma\dt\axion^{n-1}_{i'j'k'}+2\axion^{n}_{i'j'k'}-\axion^{n-1}_{i'j'k'}.
		      \end{split}
	      \end{gather}
\end{enumerate}

\section{Code Benchmark}
\label{sec:benchmark}
\begin{figure*}[htb]
	\centering
	\includegraphics[scale=0.7]{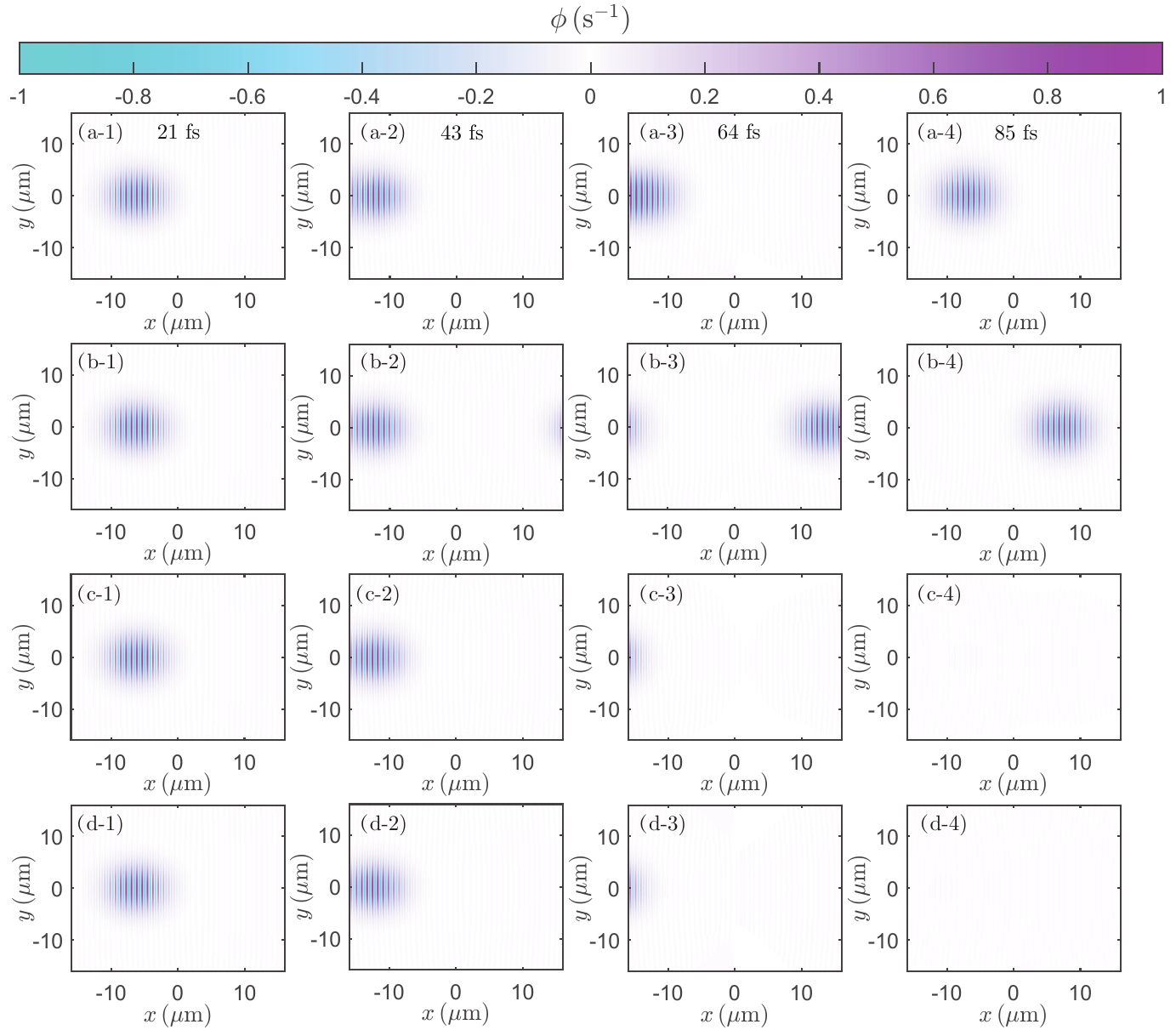}
	\caption{Temporal evolution of the axion field with different boundary conditions. Axion field distribution at different times with different boundary conditions: (a) Reflect, (b) Periodic,  (c) Outflow, (d) PML. \label{fig:result-boundary}}
\end{figure*}

After implementing the axion field and the coupling with EM fields in the code, we can self-consistently simulate the axion generation, propagation, and conversion to EM fields. We first check the different kinds of boundary conditions in the code.

\subsection{Boundary Condition Checking\label{sec:check_boundary}}

To check the boundary conditions we used in the code, we would assume a left-handed axion wave $\axion=e^{-\left(\eta^2+y^2\right)/w^2}\sin k_0\eta$ in the simulation box, where $\eta=x+ct$ is the moving coordinate, $w=5\lambda_0$ is the duration and width of the test axion pulse, and $\lambda_0=2\pi/k_0=800\,\text{nm}$ is the axion wavelength. To realize such an axion wave, what we actually do is to set the axion fields $\left.\axion\right|_{t=\frac{1}{2}\Delta t}$ and $\left.\axion\right|_{t=-\frac{1}{2}\Delta t}$ at two time steps. In this case, we considered the axion mass as 1\,meV, which is much less than the mass cooresponding to the axion wavelength set in the simulation. Since here we only check the axion boundary conditions, the coupling constant $\gagg$ does not matter. Moreover, the simulation box is set to be $[-20\lambda_0,\,20\lambda_0]\times [-20\lambda_0,\,20\lambda_0]$, and the grid number is $400\times 400$. As for the PML boundary condition, we set the layer thickness to be $N_\text{PML}=20$ grids in all directions. The maximal absorbing constant is $\sigma_\text{max}\dd t=0.2$ and the absorbing constant grows in the absorbing layer as $\sigma=\left(1-\frac{i_\text{PML}}{N_\text{PML}}\right)^3\sigma_\text{max}$, where $i_\text{PML}$ is the grid index from the boundary.

Figure~\ref{fig:result-boundary} shows the results of different boundary conditions. We can see that the axion field can be correctly reflected or propagating to another side with the corresponding reflect or periodic boundary conditions. With outflow or PML boundary conditions, the axion field can transmit out of the boundary or be absorbed by the boundary layer with negligible reflectivity. The numerical reflectivities are both  $\sim 1\%$ with the last two boundary conditions.

\subsection{Benchmark of Axion Generation\label{sec:check_generation}}
Moreover, we can further simulate the process of photon-axion and axion-photon conversions in a constant magnetic field. Since the axion wave equation contains the source term $\gagg \bm E\cdot\bm B$, a laser propagating in a constant magnetic field parallel to its polarization would continuously generate axions, and vice versa. The conversion probabilities are~\cite{arias_optimizing_2010,chou_search_2008},
\begin{gather}
	\begin{split}
		P_{\gamma\rightarrow \axion} & =P_{\axion\rightarrow \gamma}                                                                                                                            \\
		                             & \approx\frac{1}{4}\left({\gagg B_0l}\right)^2\left[\frac{\sin\left(\kappa l/2\right)}{\kappa l / 2}\right]^2,\quad\quad\left(\text{natural units}\right)
	\end{split}
\end{gather}
\begin{gather}
	\begin{split}
		\pga & =\pag                                                                                                                                                       \\
		     & \approx\frac{c}{4\mu_0\hbar}\left(\gagg B_0l\right)^2\left[\frac{\sin\left(\kappa l/2\right)}{\kappa l / 2}\right]^2,\quad\quad\left(\text{SI units}\right)
	\end{split}
\end{gather}
where $\kappa=k_1-k_0$ reflects the momentum difference between the photon and axion, $k_0,\,k_1$ are the wave vectors of photon and axion, respectively, and $l$ is the laser propagating distance inside the magnetic field.

\pgfmathsetmacro{\myscale}{0.66}
\begin{figure}[htb]
	\centering
	\includegraphics[scale=\myscale]{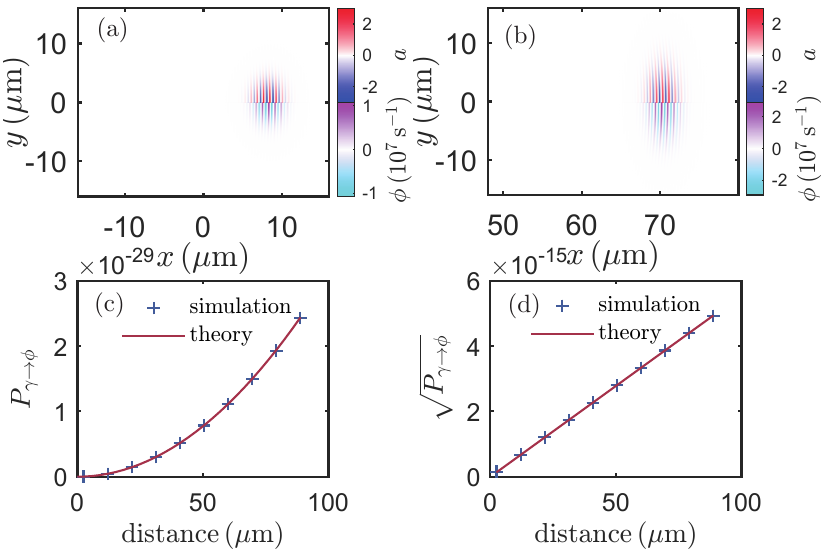}
	\caption{Laser propagation and axion generation in a constant magnetic field. Distribution of normalized laser field $a_y=\left|eE_y/(m_e\omega_0c)\right|$ (upper half) and axion field (lower half) after propagating (a) 25\,$\mu$m and (b) 89\,$\mu$m. Comparison between the conversion probability from simulation result and theoretical prediction (c) $\pga$ and (d) $\sqrt{\pga}$ during propagation.\label{fig:generation}}
\end{figure}

To simulate the process of axion generation, we consider a $p$-polarized laser with the wavelength $\lambda_0=800\,\text{nm}$ propagating in a magnetic field of $\bm B=B_\text{c}\hat{\bm y}$. The laser normalized vector potential is $a_0=\left|eE/(m_e\omega_0c)\right|=3$ and it carries Gaussian envelopes in both transverse and longitudinal directions. The width is $5\lambda_0$ and the FWHM duration is $5T_0=5\lambda_0/c$. The magnetic field is set to be 423\,T and uniform in space. The laser is propagating in $x$ direction and the simulation window is also moving at the light speed after the laser is injected into it. Moreover, here we assume the axion mass to be 1\,meV. As for the coupling constant, there are two primary models for axion-photon coupling: KSVZ (KimShifman-Vainshtein-Zakaharov)~\cite{kim_weak-interaction_1979,shifman_can_1980} and DFSZ (Dine-Fischler-Srednicki-Zhitnisky)~\cite{Zhitnitskij1980OnPS,dine_simple_1981}. Different models would give different coupling constant:
\begin{gather}
    m_a = 6.3\,\text{eV}\frac{10^6\,\text{GeV}}{f_a}, \\
    \gagg = c_\gamma \frac{2\alpha}{\pi f_a},
\end{gather}
where $c_\gamma=-0.97$ for KSVZ model and $c_\gamma=0.36$ for DFSZ model.
 Here DFSZ axion model is considered. The cooresponding coupling constant is $\gagg = 2.65\times 10^{-13}\hbar\,\text{GeV}^{-1}$. The KSVZ model would be similar.

Figure~\ref{fig:generation} shows the result of the axion generation process. The distributions of the laser electric field and axion field at two instants are shown in Fig~\ref{fig:generation}-(a,b). The laser pulse continuously generats axion and the axion field is increasing during the propagation. Meanwhile, we can also see that the laser pulse and the generated axion pulse both gradually defocus due to the finite pulse width.

We also quantitatively compared the conversion probability. The conversion probability in the simulation can be given by the total energy ratio between the axion and laser pulse in the simulation box,
\begin{gather}
	P_{\gamma\rightarrow\axion,\text{sim}} = \frac{H_{\axion}}{H_{\gamma}}=\frac{\iint \mathcal H_{\axion}\dd x\dd y}{\iint \mathcal H_{\gamma}\dd x\dd y}, \\
	\mathcal H_{\axion} = \frac{\hbar}{2c}\left(\frac{1}{c^2}\left|\partial_t\axion\right|^2+\left|\nabla\axion\right|^2+\frac{m_a^2c^2}{\hbar^2}\axion^2\right), \\
	\mathcal H_{\gamma} = \frac{1}{2}\left(\varepsilon_0 E^2+\frac{1}{\mu_0}B^2\right).
\end{gather}
The results are shown in Fig.~\ref{fig:generation}-(c,d). We can see that the conversion probabilities of theory and simulation match each other quite well, and $\sqrt{\pga}$ is linear to the propagation distance.

\subsection{Benchmark of Axion Conversion into Photons\label{sec:conversion}}

In addition, we can also simulate the process of axion conversion into photons. In this case, there is only an axion pulse initially in the simulation box and propagating in the constant magnetic field. The axion pulse is the same as that in Sec.~\ref{sec:check_boundary}, and the magnetic field is the same as that in Sec.~\ref{sec:check_generation}.

\begin{figure}[htb]
	\centering
	\includegraphics[scale=\myscale]{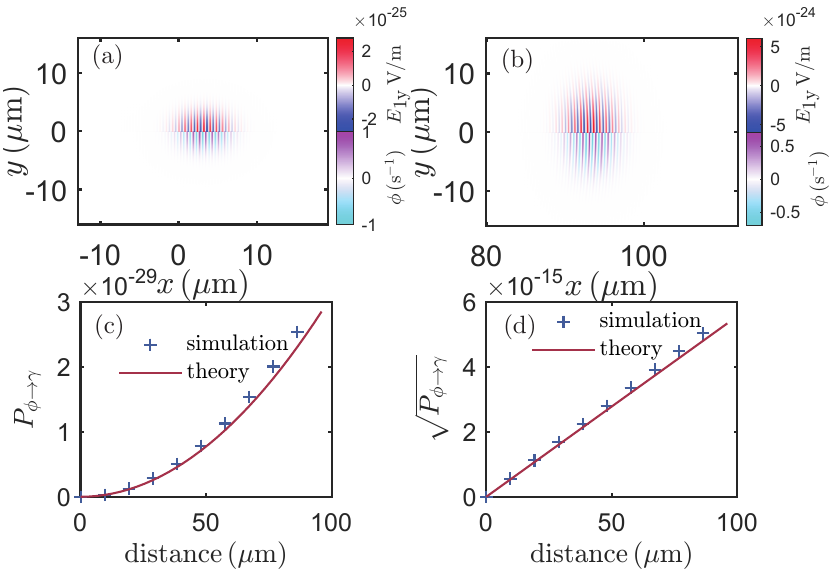}
	\caption{Axion conversion into EM fields. Distribution of perturbed laser field $E_y$ (upper half) and axion field (lower half) after propagating (a) 3\,$\mu$m and (b) 96\,$\mu$m. Comparison between the conversion probability (c) $\pag$ and (d) $\sqrt{\pag}$ during propagation.\label{fig:conversion}}
\end{figure}

As for this case, the conversion probability in the simulation is also given by the energy ratio,
\begin{gather}
	P_{\axion\rightarrow\gamma,\text{sim}} = \frac{H_{\gamma}}{H_{\axion}}=\frac{\iint \mathcal H_{\gamma}\dd x\dd y}{\iint \mathcal H_{\axion}\dd x\dd y},
\end{gather}
where the energy density $\mathcal H_\gamma$ of EM fields is given by the first order perturbation $\bm E_1,\,\bm B_1$. The results are also similar to those of photon conversion into axions, as shown in Fig.~\ref{fig:conversion}. The laser pulse is continuously generated by the axion pulse, and both two pulses gradually defocus during propagation. The conversion probability obtained from the simulation also matches well with the theory.

\subsection{Benchmark of FPS Method}
As described in Sec.~\ref{sec:perturbation}, we have used the FPS method to solve the problem of floating-point error when there is a significant strength difference between the original and derivative fields. Such a method can also be checked through an extremely weak laser pulse propagating in a magnetized plasma. We consider a laser pulse with $\lambda_0=800\,\text{nm}$ and a plasma with a density $n_0=3\times 10^{-3}n_c$, where $n_c=\varepsilon_0 m_e\omega_0^2/e^2$ is the critical plasma density. The external magnetic field is set to be $B_\text{c}=m_e\omega_p/e$ to magnetize the plasma in $z$ direction, where $\omega_p^2=n_0e^2/(\varepsilon_0m_0)$ is the plasma frequency. The laser pulse is sufficiently weak and the laser magnetic field is $10^{-20}B_\text{c}$. The simulation box is $[-20\lambda_0,\,20\lambda_0]\times[-50\lambda_0,\,50\lambda_0]$, and the space grid number is $12000\times 100$. Higher spatial accuracy was adopted in $ x$ direction, to reduce the impact of numerical dispersion. Such a laser pulse would be overwhelmed by the background magnetic if it is calculated directly. However, its propagation can be correctly calculated though the FPS method.

\begin{figure}[htb]
	\centering
	\includegraphics[scale=\myscale]{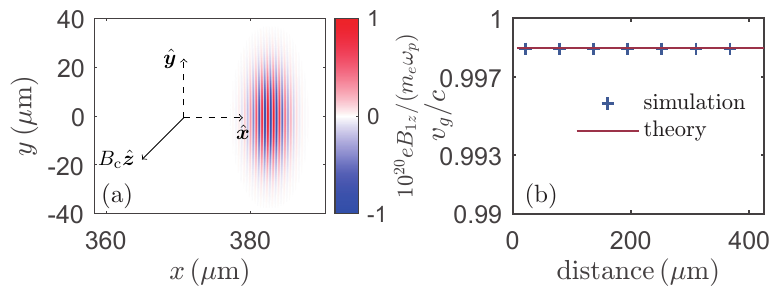}
	\caption{Weak laser propagation in a transversely magnetized plasma. (a) Distribution of the laser magnetic field. (b) Comparison between the laser group velocity from the simulation and theory.\label{fig:perturbation}}
\end{figure}

The refractivity $\eta$ of a laser in a transversely magnetized plasma is given by~\cite{chen_introduction_2016}:
\begin{gather}
	\eta^2=1-\frac{\omega_p^2}{\omega^2}\frac{\omega^2-\omega_p^2}{\omega^2-\omega_p^2-\omega_c^2},
\end{gather}
where $\omega_c={eB_\text{c}}/{m_e}$ is the electron cyclotron frequency in the magnetic field, and the magnetic field is perpendicular to the laser polarization. The the laser group velocity can be given by:
\begin{gather}
	v_g=\frac{c\eta}{\eta^2+\omega^2\frac{\dd\left(\eta^2\right)}{\dd\left(\omega^2\right)}}.
\end{gather}
On the other hand, the laser group velocity in the simulation can be given by its centroid trajectory, and it is calculated by:
\begin{gather}
	x_\text{laser}=\frac{\iint x B_{1z}^2\dd x\dd y}{\iint B_{1z}^2\dd x\dd y}.
\end{gather}

The results of weak laser propagation are shown in Fig.~\ref{fig:perturbation}. Figure~\ref{fig:perturbation} (a) shows that the sufficiently weak laser pulse can stably propagate in the magnetized plasma, rather than numerically overwhelmed by the background magnetic field. Meanwhile, as Fig.~\ref{fig:perturbation} (b) shows, the laser group velocity from the simulation matches well with the theoretical results. In fact, the FPS method can not only solve the EM modulation from the axion fields, but also correctly handling weak lasers in strong background fields.

\section{Summary}
\label{sec-conclusion}

In summary, we have proposed and benchmarked a method to model the axion field and the coupling with electromagnetic fields in the PIC code. The axion wave equation and modified Maxwell equations are considered, while the modulation of the axion field to EM fields is treated in first-order perturbation. Different boundary conditions are realized in the code. The conversion processes of axions into photons and photons into axions are checked as benchmarks of the code. The propagation of a weak laser in a strongly magnetized plasma is also checked to illustrate the validity of our FPS method. We hope such a tool can help researchers develop novel axion generation and detection schemes based on laser-plasma interactions and provide a better understanding of the astrophysical processes in which axions participate.

%%%% Acknowledgments %%%%%%%%
\section*{Acknowledgments}
\label{sec-acknowledgement}

The computations in this paper were run on the $\pi$ 2.0 cluster supported by the Center for High Performance Computing at Shanghai Jiao Tong University. This work was supported by the National Natural Science Foundation of China (No. 12225505 and No. 11991074).

%%%% Bibliography  %%%%%%%%%%
\bibliography{program_axion}
%\begin{thebibliography}{99}
%\bibitem{Berger}M. J. Berger and P. Collela, Local adaptive mesh refinement
%for shock hydrodynamics,
%J. Comput. Phys., 82 (1989), 62-84.
%\bibitem{deBoor}C. de Boor,  Good Approximation By Splines With Variable Knots II, in Springer Lecture
% Notes Series 363, Springer-Verlag, Berlin, 1973.
%\bibitem{TanTZ} Z. J. Tan, T. Tang and Z. R. Zhang, A simple moving mesh method for one- and
%two-dimensional phase-field equations, J. Comput. Appl. Math., to appear.
%\bibitem{Toro}E. F. Toro, Riemann Solvers and Numerical Methods for Fluid Dynamics,
%Springer-Verlag Berlin Heidelbert, 1999.
%\end{thebibliography}

\end{document}